# Deformation of Bacterial Cell Membranes by Action of Metal Surface under Plasmon Resonance Condition

Taras Vasyliev[1*] , Saulius Juodkazis[2], and Valeri Lozovski[1]

[1]Educational Scientific Institute of High Technologies, Taras Shevchenko National University of Kyiv, 4-g Hlushkova Avenue, Kyiv 03022, Ukraine
[2]Optical Sciences Centre, Australian Research Council (ARC) Industrial Transformation Training Centre in Surface Engineering for Advanced Materials (SEAM), Swinburne University of Technology, Hawthorn, VIC 3122, Australiay
*taras.vasyliev@knu.ua

**ABSTRACT**. This paper is devoted to studies of the mechanical deformation of the S. aureus cell wall. The bacterium is modelled as a thin elastic membrane containing cytoplasm, which is treated as an incompressible fluid. Deformation occurs via Van der Waals interactions between the bacterium and a solid metallic surface, both with and without the influence of surface plasmon resonance (SPR). Our modelling results indicate that the excitation of surface plasmons significantly increases the effective interaction area between the bacterial membrane and the nanostructured surface. The elastic and dielectric properties of the bacterium's components are uninvestigated. Therefore, theoretical calculations are performed in wide, physically meaningful ranges. Thus, the results of studies give only a qualitative estimation. However, they are novel and, with further experiments, can solve the inverse problem of obtaining physical properties. The paper highlights the potential of SPR to enhance antibacterial strategies, inspiring further research and innovation.

## I. INTRODUCTION.

It is known that the interaction of bacteria/viruses with a metal surface or nanoparticles leads to their loss of infectious activity [1-5]. However, the mechanisms underlying the antimicrobial/antiviral action of these interactions are not fully elucidated. In the experimental work [1], it was shown that, upon interaction with metal nanoparticles, membrane tension increases due to adsorption. Mechanical deformation leads to rupture of the cell membrane and the death of bacteria such as Gram-positive Staphylococcus aureus and Gram-negative Pseudomonas aeruginosa. The significance of membrane rupture was demonstrated by the direct observation of tension and compression of the bacterial membrane induced by gold nanoparticles. The results, obtained using a specially designed microfluidic device that recorded membrane deformation and lipid bilayer disruption, provide critical insights into nanoparticle-induced bacterial cell death. Electron microscopy methods characterised the interaction between nanoparticles and the bacterial membrane, supporting the proposed membrane-tension-induced (mechanical) killing mechanism.

Many studies investigate the deformation of soft nanoparticles, particularly biological cells. The studies reported in [6-10] were devoted to modelling the mechanical properties of a soft particle consisting of a thin elastic membrane filled with an incompressible liquid (cytoplasm). These works are based on the well-known theory of cell deformation within the framework of nonlinear elasticity [11-13]. The mechanical properties of a nanoparticle located in an electrolyte were considered in [7]. Here, it was assumed that nanoparticles have a uniform electrical charge distributed over their surfaces and adhere to the surface of an oppositely charged solid substrate via electrostatic attractive forces. In other works [8,9], it was assumed that adhesion occurs through the interaction of cell receptors with ligands on the surface of a solid. The case in which bacteria attach to a solid surface due to mechanical forces was considered in [10,11]. Note that in all cases, it was considered that the nanoparticle (bacterium) is capable of undergoing significant nonlinear deformations. The model used in these works allows for the study of the relationship between the forces and changes in the shape of the cell membrane. The model predicts the membrane response upon contact with the substrate and provides useful insights into how varying parameters, such as the distance between the nanoparticle and the surface, the membrane shear modulus, etc can modulate adhesion. The model allows for the study of membrane deformations under the action of significant mechanical forces and of cell receptors with ligands on the surface of a solid. Note that the aforementioned models do not account for the deformation of the bacterial cell membrane when Van der Waals interactions occur between the bacteria and the solid surface. Moreover, the model does not account for membrane bending, so it is valid only for thin membranes whose thickness is much less than the bacterium's radius. Given that the thickness of a bacterial membrane is typically two orders of magnitude less than the bacteria's dimensions, the model used is valid.

In the present work, a bacterium is modelled as a thin elastic membrane with its cytoplasm, treated as an incompressible fluid, inside. The deformation of the bacterium and the pressure of the fluid inside it, resulting from its interaction with a metal surface via nonlinear Van der Waals interactions, particularly under conditions of surface plasmon resonance, are studied. This focus aims to deepen understanding of membrane deformation mechanisms driven by surface forces, with relevance to microbiology and nanotechnology.

## II. STATEMENT OF THE PROBLEM

Consider the system consisting of a spherical bacterium (for example, S. aureus) located at the surface of a solid. To calculate the deformation of the bacteria due to their interaction with a metallic surface under plasmon-resonant conditions, we consider a spherical membrane of radius R in its base configuration above the plane metal surface (Fig. 1).

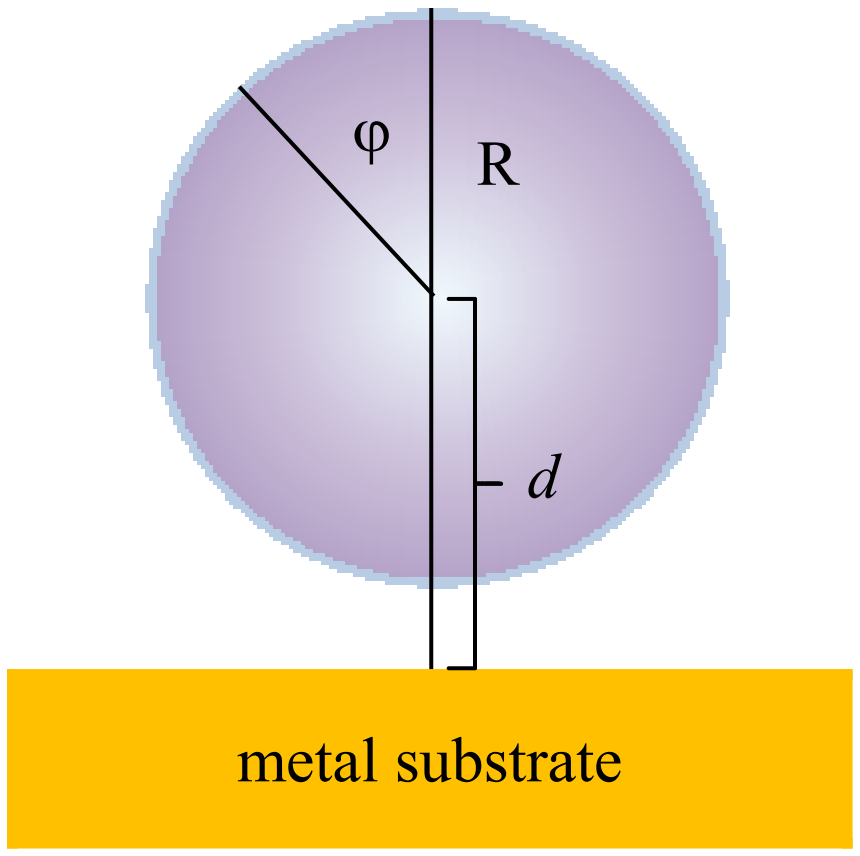

FIG 1. The sketch of the model.

It is supposed that, because the membrane is very thin (about a few nanometers in thickness) [6-11], it has no bending stiffness. However, the membrane is also ductile and can withstand large nonlinear deformations. The intrinsic volume of a sphere is filled with incompressible fluid. Suppose the bacterium adsorbs to the metal surface via the van der Waals mechanism [14,15], with nonlinear polarizability taken into account [16]. The interaction between the particle and the surface of a solid was studied within the framework of the effective susceptibility concept [17], including the case of surface plasmon excitation on the metal surface (under resonant conditions). The adsorption potential was calculated. It was shown, particularly, that excitation of surface plasmons leads to a sharp (approximately an order of magnitude) increase in the binding energy. Further, the membrane interacts with a surface, but the cytoplasm inside the membrane only responds to shell deformation. The adsorption potential consequence of Ref. [16] – the dependence of energy interaction between the particle and a surface – in the cases both under surface plasmon resonance and without excitation, the surface plasmon can be expressed as

$$U(d,\omega) = -\int_V P_j E_j d\mathbf{R} - \delta U_\infty \,, \qquad (1)$$

where $d$ is a distance to the metal surface, $V$ is the volume of a particle, $E_j$ is a local field, and $P_j$ is an electric dipole momentum density. Integration is over the particle volume (where the dipole momentum is nonzero). The last term in Eq.(1) is the energy of the system when the particle is moved from the surface to an infinite distance

$$\delta U_\infty = -\int_V P_j E_j d\mathbf{R}\Big|_{d\to\infty} \,. \qquad (2)$$

Thus, the adsorption potential has an general form

$$U(d) = \int_0^{+\infty} U(d,\omega) f(\omega) d\omega \,, \qquad (3)$$

with $f(\omega) = \left[\exp(h\omega / kT) - 1\right]^{-1}$ Bose function of distribution for virtual photons. As a result, the adsorption potential views as [16]

$$U(d) = A\left(\frac{M}{d^3} - \frac{N}{d^6}\right), \qquad (4)$$

were $M$ and $N$ are the coefficients defined by electrodynamic properties of the bacterium, nutrient medium and metal surface

$$M = \frac{5}{2\pi\varepsilon_0\alpha\beta}\frac{\varepsilon_s - \varepsilon_m}{\varepsilon_s + \varepsilon_m}, \quad N = \frac{99}{32\pi^2\varepsilon_0\beta}\left(\frac{\varepsilon_s - \varepsilon_m}{\varepsilon_s + \varepsilon_m}\right)^2, \tag{5}$$

with $\varepsilon_s, \varepsilon_m$ dielectric constants of nutrient medium and a metal, $\alpha$ and $\beta$ linear and nonlinear polarizabilities of bacterium, respectively:

$$\alpha = 3V\frac{\varepsilon_b - \varepsilon_m}{\varepsilon_b + 2\varepsilon_m}, \quad \beta = \frac{\gamma}{\varepsilon_0^3\alpha^4}, \tag{6}$$

where $\varepsilon_b$ is the dielectric constant of the bacterium cell cytoplasm, $\gamma$ and $\varepsilon_0$ is the third order hyper-polirazibility and the vacuum permittivity, respectively. Parameter $A$, introduced into equation (4), is responsible for the presence or absence of plasmon resonance on the metal surface.

As a result, the force of interaction is

$$F_{ns}(d) = -\frac{\partial U(d)}{\partial d} = 3A\left(\frac{M}{d^4} - \frac{2N}{d^7}\right). \tag{7}$$

## III. BACTERIAL CELL MEMBRANE DEFORMATION ENERGY

The model based on the nonlinear theory of elasticity is used to describe the deformation of the bacterial membrane, inside which the incompressible fluid (cytoplasm) is located [7-11]. The model's main features stem from the following circumstances.

Interaction with the metal surface leads to deformation of the bacterial membrane. The deformation energy determines the stresses within the membrane material. Phospholipid molecules, the primary component of the membrane, are held together by noncovalent interactions. Consequently, the characteristic behaviour of the membrane during stretching and torsion is closely related to the behaviour of the internal cytoplasm. Therefore, stretching the bacterial membrane is an energetically costly process. However, the membrane has rather small resistance to distortion. Thus, the deformation energy accounts for the resistance of fluid-like membranes to surface expansion; however, it is not sensitive to membrane distortions. Thus, one can use it for establishing the strain energy density function of the bacterial membrane [9,12]

$$W(\lambda,\mu) = \frac{G}{2}(J-1)^2, \quad J = \lambda\mu, \tag{8}$$

where $G$ is the membrane shear modulus, $J$ is the areal dilation of the membrane surface, and $\lambda$ and $\mu$ are the relative meridional and circumferential deformations, respectively.

The choice of the strain energy function is due to physical considerations. Without the strains ( $J=1$ ), the strain energy must attain its minimum (zero) value because, at the base configuration, the strain energy must be minimal and the stresses in the membrane disappear. Using Eq.(8), one obtains the necessary and sufficient conditions for an extremum

$$\frac{dW(J=1)}{dJ} = G(J-1)\Big\|_{J=1} = 0, \tag{9}$$

$$\frac{d^2W(J=1)}{dJ^2} = G > 0. \tag{10}$$

The next physical restriction on the choice of the deformation function arises from the condition that the stress as a function of the deformation tensor must monotonically increase. It requires the convexity of the strain energy function under arbitrary deformation. This condition is fulfilled because

$$\frac{d^2W(J)}{dJ^2} = G > 0. \quad (11)$$

Thus, Eqs. (9-11) will be starting equations for obtaining the ground state of the bacterium interacting with the surface of a solid.

## IV. GRADIENT OF DEFORMATION TENSOR AND CAUCHY TENSOR

The bacterial membrane in the model under consideration is essentially a two-dimensional object in three-dimensional space. For the description of shifting the points at the membrane, it is convenient to use the standard spherical angle coordinates (ϕ,θ). Thus, the points of a surface of a membrane can be described by a radius vector

$$\mathbf{X}(\phi,\theta) = R\,\mathbf{e}_R(\phi,\theta)\ . \quad (12)$$

Under deformation, the point on the surface shifts and will be characterised by a radius vector with coordinates $u(\phi), h(\phi)$ (Fig.2)

$$\mathbf{X}(\phi,\theta) = u\,(\phi)\mathbf{e}_u(\theta) + h\,(\phi)\mathbf{e}_h\ . \quad (13)$$

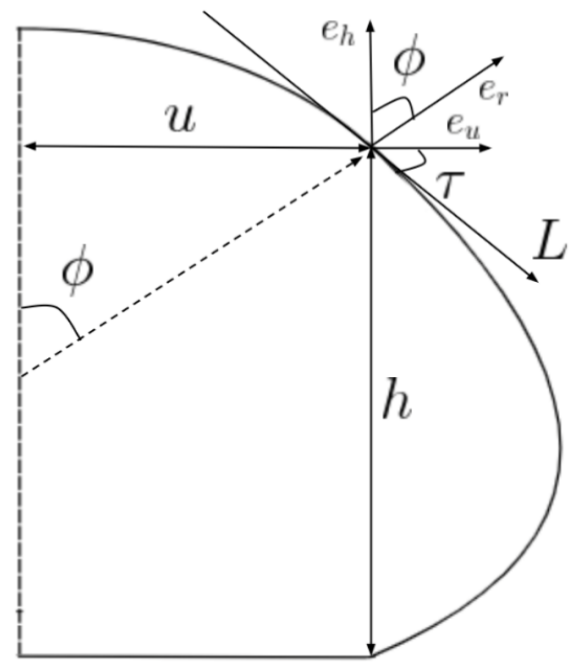

FIG 2. Cylindrical coordinates of deformed bacterium.

Note there is no dependence on the azimuth angle, which reflects the axis-symmetry in bacterium deformation. In the frame of nonlinear elasticity theory, the differential connection is between the values given by Eqs. (12) and (13) [9]

$$d\mathbf{x}(\phi,\theta) = \hat{\mathbf{F}}(\phi,\theta)d\mathbf{X}(\phi,\theta)\ , \quad (14)$$

with the tensor $\hat{\mathbf{F}}(\phi,\theta)$ of the gradient of deformation. The mathematical procedure for establishing $\hat{\mathbf{F}}(\phi,\theta)$ was performed in [7]. The result of the calculations is

$$\hat{\mathbf{F}}(\varphi,\theta) = \lambda(\varphi)\mathbf{l}(\varphi,\theta) \otimes \mathbf{L}(\varphi,\theta) + \mu(\varphi)\mathbf{m}(\theta) \otimes \mathbf{M}(\theta)\ , \quad (15)$$

where $\lambda(\phi)$ and $\mu(\phi)$ are the meridional and the circumferential relative deformations, respectively

$$\lambda(\phi) = \frac{\sqrt{(u'(\phi))^2 + (h'(\phi))^2}}{R}, \quad \mu(\phi) = \frac{u(\phi)}{R\sin\phi}\ , \quad (16)$$

and

$$\mathbf{l}(\phi,\theta)=\frac{u'(\phi)\mathbf{e}_u(\theta)+h'(\phi)\mathbf{e}_h}{\lambda(\phi)R},\ \mathbf{L}(\phi,\theta)=\mathbf{e}_\phi(\phi,\theta),$$
$$\mathbf{m}(\theta)=\mathbf{e}_\theta(\theta),\quad \mathbf{M}(\theta)=\mathbf{E}_\theta(\theta). \tag{17}$$

As for the Cauchy stress tensor in the membrane, it is determined from the strain energy function $W$ discussed in the previous section. In the case under consideration, it has the form

$$\hat{\mathbf{T}}(\lambda,\mu)=G(J-1)(\hat{\mathbf{I}}-\mathbf{n}\otimes\mathbf{n})\ . \tag{18}$$

## V. MATHEMATICAL DESCRIPTION OF DEFORMATION OF BACTERIAL MEMBRANE

Suppose the bacterial deformation under the forces described in part 1 is slow. The equation of equilibrium has the form [8]

$$\mathrm{div}\hat{\mathbf{P}}+J\mathbf{F}=0\ , \tag{19}$$

where $\hat{\mathbf{P}}$ is the first tensor Piola

$$\hat{\mathbf{P}}=J\hat{\mathbf{T}}\hat{\mathbf{F}}^{-T},$$
$$\hat{\mathbf{F}}^{-T}=\lambda^{-1}(\varphi)\mathbf{l}(\varphi,\theta)\otimes\mathbf{L}(\varphi,\theta)+\mu^{-1}(\varphi)\mathbf{m}(\theta)\otimes\mathbf{M}(\theta), \tag{20}$$

with $\mathbf{F}$ the force acting to the unit square of membrane

$$\mathbf{F}=\left(p_f(d)-p_c\right)\mathbf{n}+\mathbf{F}_{ns}\delta\ . \tag{21}$$

Here, $p_f$ and $p_c$ are the hydrostatic pressure of a liquid and the contact pressure on the external surface of the membrane, respectively, δ is the thickness of a bacterial shell (note, $\delta\approx 28$ nm for Gram-positive Staphylococcus aureus). The pressure of a liquid depends on the distance to the metal surface because of the cytoplasm interacting with the surface

$$p'_f(d)=-F_{ns}\ . \tag{22}$$

From this equation, one finds the dependence of cytoplasmic pressure on the distance from the metal surface

$$p_f(d)=p_f(h_0)+U(d)\ , \tag{23}$$

where $p_f(h_0)=p_{f0}$ is the pressure in the cytoplasm at a distance $h_0$ from the metal surface (in the closest point to the surface), and $U(d)$ is the interaction potential given by Eq. (4).

Substituting the explicit expressions of the Cauchy tensor and the tensor of gradient of deformation into the Piola tensor and calculating the divergence, one obtains

$$\mathrm{div}\hat{\mathbf{P}}=\left[\frac{\partial^2 W}{\partial\lambda^2}\lambda'+\frac{\partial^2 W}{\partial\lambda\partial\mu}\mu'+\frac{\partial W}{\partial\lambda}\cot\varphi\right]\frac{\mathbf{l}}{R}+$$
$$+\frac{\partial W}{\partial\lambda}\frac{\mathbf{l}'}{R}-\frac{\partial W}{\partial\mu}\frac{1}{R\sin\varphi}\mathbf{e}_u. \tag{24}$$

This result can be substituted into Eq. (19). Further, one projects it onto the tangential and normal directions to the surface of the deformed membrane. Thus, one obtains two nonlinear differential equations for meridional relative deformation and the angle between the horizontal direction and the tangential line to the deformed surface of a membrane

$$\lambda' = \frac{\left(W_{\mu} - \lambda W_{\lambda\mu}\right)\cos\tau - \left(W_{\lambda} - \mu W_{\lambda\mu}\right)\cos\varphi}{W_{\lambda\lambda}\sin\varphi} - $$
$$- \frac{RJF_{ns}\delta\sin\tau}{W_{\lambda\lambda}}; \tag{25}$$
$$\tau' = \frac{RJ\left(p_{f0} + U - p_c - F_{ns}\delta\cos\tau\right)}{W_{\lambda}} - \frac{W_{\mu}}{W_{\lambda}}\frac{\sin\tau}{\sin\varphi}.$$

This system of equations should be supplemented by a connection between the coordinates of points $u(\varphi)$ and $h(\varphi)$ of the deformed membrane, and the relative meridian deformation of a membrane $\lambda(\varphi)$ and the angle $\tau(\varphi)$, which is caused by relationships (16) and (17)

$$\begin{aligned} u' &= \lambda R\cos\tau, \\ h' &= -\lambda R\sin\tau. \end{aligned} \tag{26}$$

Eqs. (25) and (26) are the system of nonlinear differential equations of first order describing the deformation of the bacterial membrane. It is convenient to use dimensionless quantities. To provide dimensionlessness, let one suppose that a spherical bacterium has a radius R when the density of the inner liquid inside the bacterium is equal to the density of the external nutrient medium. However, in reality, the liquid inside is denser, so the bacterial shell is somewhat taut before interacting with the metal surface, and the bacterium itself is slightly inflated into a sphere of radius $r_0 \approx 1.01R\div1.1R$. Let us introduce the dimensionless diameter of the bacterium before interacting with the metal surface

$$\bar{D}_0 = 2r_0 / R\ . \tag{27}$$

All length dimension values in relations are conveniently normalised to $2r_0$, which is the diameter of the unperturbed bacterium, such that

$$\bar{u}(\varphi) = \frac{u(\varphi)}{2r_0}, \quad \bar{h}(\varphi) = \frac{h(\varphi)}{2r_0}\ . \tag{28}$$

Despite the meridional and circular relative deformations λ and μ being naturally dimensionless by definition, they can now be represented using dimensionless coordinates

$$\lambda(\varphi) = \bar{D}_0\sqrt{\left(\bar{u}'\right)^2 + \left(\bar{h}'\right)^2},\ \mu(\varphi) = \frac{\bar{D}_0\bar{u}(\varphi)}{\sin\varphi}\ . \tag{29}$$

The strain energy function and, accordingly, all its partial derivatives can be made dimensionless by dividing by the shear modulus, i.e.

$$\bar{W}(\lambda,\mu) = \frac{W(\lambda,\mu)}{G}\ . \tag{30}$$

Using expressions (27)-(30) we obtain the dimensionless system

$$
\begin{cases}
\bar{u}' = \dfrac{1}{\bar{D}_0}\lambda\cos\tau;\\
\bar{h}' = -\dfrac{1}{\bar{D}_0}\lambda\sin\tau;\\
\lambda' = \dfrac{\left(\bar{W}_\mu - \lambda\bar{W}_{\lambda\mu}\right)\cos\tau - \left(\bar{W}_\lambda - \mu\bar{W}_{\lambda\mu}\right)\cos\varphi}{\bar{W}_{\lambda\lambda}\sin\varphi} -\\
\quad -\dfrac{J\bar{F}_{ns}\bar{\delta}\sin\tau}{\bar{W}_{\lambda\lambda}};\\
\tau' = \dfrac{J\left(\left(\bar{D}_0^2-4\right)(\bar{p}_{f0}+\bar{U}-\bar{p}_c)/\bar{D}_0 - \bar{F}_{ns}\bar{\delta}\cos\tau\right)}{\bar{W}_\lambda} -\\
\quad -\dfrac{\bar{W}_\mu}{\bar{W}_\lambda}\dfrac{\sin\tau}{\sin\varphi}.
\end{cases}
\tag{31}
$$

The system (31) can be simplified. From the first, third and fourth equations, we obtain

$$
\begin{aligned}
&\left(\bar{W}_\lambda\sin\varphi\cos\tau\right)' =\\
&= \bar{W}_\mu - J\frac{\bar{D}_0^2-4}{\bar{D}_0}\left[\bar{p}_{f_0}+U-\bar{p}_c\right]\sin\tau\sin\varphi;\\
&\left(\bar{W}_\lambda\sin\varphi\sin\tau\right)' =\\
&= J\sin\varphi\left(\frac{\bar{D}_0^2-4}{\bar{D}_0}\left[\bar{p}_{f_0}+U-\bar{p}_c\right]\cos\tau - \bar{F}_{ns}\bar{\delta}\right).
\end{aligned}
$$

This leads to

$$
\begin{aligned}
&\frac{dJ}{d\bar{h}}\cos\tau + (J-1)\sin\tau\left[\frac{1}{\bar{u}} - \frac{d\tau}{d\bar{h}}\right] -\\
&\quad -\left(\bar{D}_0^2-4\right)\left[\bar{p}_{f_0}+U\right] = 0;\\
&\frac{d\tau}{d\bar{h}} = \frac{1}{\bar{u}} - \frac{\left(\bar{D}_0^2-4\right)\left[\bar{p}_{f_0}+U\right] - \bar{F}_{ns}\bar{\delta}\bar{D}_0\cos\tau}{(J-1)\sin\tau}.
\end{aligned}
$$

Substitution the second equation into the first simplifies the first one leads to

$$
\begin{aligned}
&\frac{dJ}{d\bar{h}}\cos\tau - \bar{F}_{ns}\bar{\delta}\bar{D}_0\cos\tau = 0 \quad\Rightarrow\\
&\Rightarrow\quad \frac{dJ}{d\bar{h}} = \bar{F}_{ns}(\bar{h}_0+\bar{h})\bar{\delta}\bar{D}_0\,.
\end{aligned}
$$

Further integration gives us

$$
\begin{aligned}
&J(\bar{h}) = \lambda(\bar{h})\mu(\bar{h}) =\\
&= \bar{\rho}_m\bar{\delta}\bar{D}_0\int_{\bar{h}_T}^{\bar{h}} F_{ns}(\bar{h}_0+x)dx + \lambda\left(\bar{h}_T\right)\mu\left(\bar{h}_T\right) =\\
&= \bar{\rho}_m\bar{\delta}\bar{D}_0\left[U_{ns}(\bar{h}_0+\bar{h}_T) - U_{ns}(\bar{h}_0+\bar{h})\right] + \lambda^2\left(\bar{h}_T\right).
\end{aligned}
$$

Here, we used that in the polar point $\bar{h}=\bar{h}_T$ the meridianial and circumferential relative deformations are equal $\lambda\left(\bar{h}_T\right)\equiv\mu\left(\bar{h}_T\right)$, $\bar{h}_0$ is the distance between the metal surface and the bacterium. Finally, one can find a relation between $\lambda(\bar{h})$ and $\bar{u}(\bar{h}(\varphi))$

$$
\begin{aligned}
&\lambda(\bar{h}) =\\
&= \frac{\left\{\bar{\delta}\bar{D}_0\left[U_{ns}(\bar{h}_0+\bar{h}_T) - U_{ns}(\bar{h}_0+\bar{h})\right] + \lambda^2\left(\bar{h}_T\right)\right\}\sin\varphi}{\bar{u}(\bar{h}(\varphi))\bar{D}_0}.
\end{aligned}
\tag{32}
$$

Therefore, the system (31) transforms to a simplified system of three equations for $\bar{u}(\varphi),\ \bar{h}(\varphi),\ \tau(\varphi)$ .

Since the repulsive forces prevail in the zone closest to the metal surface, the bacterium can approach the metal surface at a distance $\bar{h}_0$. As a result of the repulsive force, a contact region arises at $\varphi > \varphi_c$ , where $\tau' = 0$ and $\bar{h}' = 0$ , and the system (31) reduces to two equations

$$\begin{cases} \bar{u}' = -\dfrac{1}{\bar{D}_0}\lambda; \\ \lambda' = -\dfrac{\left(\bar{W}_\mu - \lambda \bar{W}_{\lambda\mu}\right) + \left(\bar{W}_\lambda - \mu \bar{W}_{\lambda\mu}\right)\cos\varphi}{\bar{W}_{\lambda\lambda}\sin\varphi}. \end{cases} \tag{33}$$

The system (33) can be solved exactly. The second equation of (33) is simplified

$$\begin{aligned} &\left(\bar{W}_\lambda \sin\varphi\right)' + \bar{W}_\mu = 0 \quad \Leftrightarrow \\ &\Leftrightarrow \left(\mu \bar{W}_J \sin\varphi\right)' + \lambda \bar{W}_J = 0 \quad \Rightarrow \ . \\ &\Rightarrow \ J = C_1 = const \end{aligned}$$

Therefore,

$$\lambda = \frac{C_1}{\mu} = \frac{C_1 \sin\varphi}{\bar{D}_0 \bar{u}} \ . \tag{34}$$

Substitution (34) in the first equation of the system (32) gives separable differential equation

$$\bar{u}' = -\frac{C_1 \sin\varphi}{\bar{D}_0^2 \bar{u}} \ . \tag{35}$$

The general integral of (35) is

$$\bar{u} = \sqrt{2\frac{C_1}{\bar{D}_0^2}\cos\varphi + C_2} \ . \tag{36}$$

Solution of Eqs. (31) and relations (32), (34) and (36) with appropriate border conditions can give us the change of bacterial shape when bacteria approach the surface.

## VI. BOUNDARY CONDITIONS AND METHOD OF SOLUTION OF THE EQUATIONS

The system of differential equations Eqs. (31) in the noncontact domain were solved by taking into account the next four boundary conditions

$$\bar{u}(0) = 0; \quad \bar{h}(\varphi_c) = 0; \ \tau(0) = 0; \ \tau(\varphi_c) = \pi \ . \tag{37}$$

The system of equations (32), we supplement with boundary conditions

$$\bar{u}(\pi) = 0; \quad \bar{u}(\varphi_c) = a \ , \tag{38}$$

where parameter a is defined during the solution of the boundary problem [Eqs.(31)-(32)].
Substitution (36) into (34) gives for $\bar{u}(\varphi), \lambda(\varphi),\ \varphi_c \le \varphi \le \pi$

$$\bar{u} = a\frac{\cos\frac{\varphi}{2}}{\cos\frac{\varphi_c}{2}}, \qquad \lambda(\varphi) = \frac{a\bar{D}_0 \sin\frac{\varphi}{2}}{2\cos\frac{\varphi_c}{2}} \ . \tag{39}$$

Since the boundary conditions are specified at the initial or end points of the domain, boundary value problems can be solved numerically using the shooting method [10,11]. In this case, the boundary-value problem is transformed into four Cauchy problems with initial conditions at the endpoints. Of course, there are not enough initial conditions, so the unknown initial values of the quantities are given arbitrarily. In detail, this means that the domain of integration of the system (31) by coordinate is divided into $2n$+$2m$ points

$$\left\{\varphi_1 = 0,\ldots,\ \varphi_n = \frac{\pi}{4},\ldots,\ \varphi_{n+m} = \frac{\pi}{2},\ldots,\right.$$
$$\left.\varphi_{2n+m} = \frac{3\pi}{4},\ldots,\varphi_{2n+m+i} = \varphi_c,\ldots,\varphi_{2n+2m} = \pi\right\}.$$

In the intervals $\left[0;\frac{\pi}{4}\right]$ and $\left[\frac{\pi}{2};\frac{3\pi}{4}\right]$, the straightforward direction of the fourth-order Runge-Kutta method was used. At the same time, in the intervals $\left[\frac{\pi}{4};\frac{\pi}{2}\right]$ and $\left[\frac{3\pi}{4};\pi\right]$, the inverse direction was applied.

Further, the known values of Eq. (31) as on a top pole $\varphi = 0$ as on a bottom pole $\varphi = \pi$ are considered as the initial values of the first and the fourth Cauchy problems. For a correct setup of the problem in the top pole $\varphi = 0$ one needs to add the conditions $\bar{h}(0) = \bar{h}_T$, $\lambda(0) = \lambda_T$ ($\bar{h}_T$, $\lambda_T$ are some randomly chosen values). Analogously, to find a solution of the fourth Cauchy problem with an initial point $\varphi = \pi$ one adds the conditions $\bar{u}(\varphi_c) = a$ with $a$ some randomly chosen value. Since (32) and (39) must be the same in the point $\varphi = \varphi_c$, therefore

$$\lambda(\varphi_c) = \frac{a\bar{D}_0 \sin\frac{\varphi_c}{2}}{2\cos\frac{\varphi_c}{2}} = \frac{\lambda^2\left(\bar{h}_T\right)\sin\varphi_c}{a\bar{D}_0} \Rightarrow$$
$$\Rightarrow \quad a = \frac{2\lambda_T \cos\frac{\varphi_c}{2}}{\bar{D}_0}.$$

For the other two Cauchy problems, one needs to choose $\bar{u}\left(\frac{\pi}{2}\right) = \bar{u}_M, \bar{h}\left(\frac{\pi}{2}\right) = \bar{h}_M, \tau\left(\frac{\pi}{2}\right) = \tau_M$.

Therefore, the seven unknown parameters were used $\bar{h}_T, \lambda_T, \bar{u}_M, \bar{h}_M, \tau_M, \varphi_c, \bar{p}_{f0}$.

To find these parameters, we have eight meaningful conditions. To prevent shell rupture, the solutions to the Cauchy problems must converge at points $\varphi_n = \pi/4$ and $\varphi_{2n+m} = 3\pi/4$ i.e. they must obey six conditions

$$\begin{aligned}&\left[\bar{u}(\varphi_n)\right] = \left[\bar{h}(\varphi_n)\right] = \left[\tau(\varphi_n)\right] = \\ &= \left[\bar{u}(\varphi_{2n+m})\right] = \left[\bar{h}(\varphi_{2n+m})\right] = \left[\tau(\varphi_{2n+m})\right] = 0\end{aligned} \tag{40}$$

with $\left[f(\varphi_i)\right] = f(\varphi_i + 0) - f(\varphi_i - 0)$. We note that the unknown in the problem is also the cytoplasmic pressure on the front surface of the bacterial shell. Initially, we set it at our discretion, but in the end, we chose it based on the liquid's incompressibility within the bacterium's membrane. Note that the incompressibility of the liquid (cytoplasm) requires a constant volume for the bacterium. The initial volume of the bacterium, associated with the reference configuration (undeformed cell), is equal to

$$V_0 = \frac{4}{3}\pi r_0^3 . \tag{41}$$

Using the determined length $d_0$=$2r_0$ as the initial diameter of the bacterium, one obtains

$$\bar{V}_0 = \frac{V_0}{(2r_0)^3} = \frac{\pi}{6}. \quad (42)\text{is}$$

The volume of a deformed bacterium is determined as [8]

$$\bar{V} = \frac{\pi}{3}\sum_i \left(\bar{u}_{i+1}^2 + \bar{u}_i\bar{u}_{i+1} + \bar{u}_i^2\right)\left(\bar{h}_i - \bar{h}_{i+1}\right), \\ \bar{u}_i = \bar{u}(\varphi_i), \quad \bar{h}_i = \bar{h}(\varphi_i). \quad (43)$$

Thus, the condition of constant volume of the bacterium is

$$\frac{\bar{V}}{\bar{V}_0} = 1. \quad (44)$$

In practice, we use the Hooke-Jeeves algorithm to find the null of the target function

$$TF(\bar{h}_T, \lambda_T, \bar{u}_M, \bar{h}_M, \tau_M, \varphi_c, \bar{p}_{f0}) = \\ = [\bar{u}(\varphi_n)]^2 + [\bar{h}(\varphi_n)]^2 + [\tau(\varphi_n)]^2 + \\ + [\bar{u}(\varphi_{2n+m})]^2 + [\bar{h}(\varphi_{2n+m})]^2 + [\tau(\varphi_{2n+m})]^2 + \left(\frac{\bar{V}}{\bar{V}_0} - 1\right)^2.$$

## VII. NUMERICAL CALCULATIONS

Physical properties of the Gram-positive bacterium S. aureus were the subject of numerous studies [18-22]. However, they were estimated in a wide range of values. For instance, the elastic properties of bacterial membranes were studied in [18] and [19]. In this work, the Gram-positive bacterium Staphylococcus aureus [20] is modelled, with a shape close to spherical. For such a bacterium, it was found that the shortened Young's modulus of the bacterial membrane

$$E^* = \frac{E}{1-\nu^2} = \frac{2G}{1-\nu}, \quad (45)$$

where $G$ is a shear modulus, and ν is Poisson's ratio. The value of $E^*$ is in the range of 8÷47 kPa [19]. Because Poisson's ratio lies in the range $-1 < \nu < 0.5$ (for real materials $0 < \nu < 0.5$), the shear modulus is in the range of 2÷47 kPa (2÷23,5 kPa). The thickness of the membrane was estimated to be close to 8nm [21]. The thicker bacterial wall (approximately 20nm) was much stronger. The Young's modulus of the wall is close to 1.5MPa [22]. Therefore, the shear modulus is in the range 0.5-0.75MPa. In our studies, we supposed the bacterium wall and the membrane can be approximated by a homogeneous shell with an average value of shear modulus

$$G_{shell} = \frac{8}{28}G_{membrane} + \frac{20}{28}G_{wall}, \quad (46)$$

In our case, the value of the shear modulus of the shell was in the range 0.36-0.55 MPa.
The dielectric properties of the bacterium S areus. wall ($\varepsilon_{wall} \approx 60$), membrane ($\varepsilon_{membr} \approx 16$) and cytoplasm ($\varepsilon_{cyt} \approx 70$) were studied in Ref. [21]. Because of the membrane's thinness, we use the dielectric constant $\varepsilon_b = 65$ for the whole bacterium. The dielectric constant of the gold surface was taken $\varepsilon_m \approx -15$. The dielectric properties of the physical solution was $\varepsilon_s \approx 80$. The bacterium size was $d_0 = 1\ \mu m$. For such values of physical parameters, the value of equilibrium distance between the bacterium and the gold surface was near $h_0 \sim 0.9d_0$. The linear polarizability $\alpha \approx 4.19 \cdot 10^{-18} m^3$. The second-order hyperpolarizability γ for a bacterium is unknown, but for other organic materials it is close to $\gamma \sim 10^{-58} - 10^{-54} C^4 m^2 J^{-3}$ [23]. Here, the upper limit of the second-order hyperelasticity is huge and exceeds by two orders the hyperpolarizability of polyene oligomers. The value of $\beta = \frac{\gamma}{\varepsilon_0^3 \alpha^4}$ in (6) is $\beta \sim 5 \cdot 10^{44} - 5 \cdot 10^{48} C^{-2} m^{-5}$. Further evaluations yield values of coefficients *M* and *N*: $M \sim 10^{-20} - 10^{-16} J$, $N \sim 1.363 \cdot 10^{-38} - 1.363 \cdot 10^{-34} J \cdot m^3$. It must be noted that in dimensionless shape, the interaction potential (6) is

$$U(\bar{h}) = A\xi \left( \frac{1}{\left(\bar{h} + \bar{h}_0\right)^3} - \frac{1.363}{\left(\bar{h} + \bar{h}_0\right)^6} \right). \tag{47}$$

Here $\xi$ is a dimensionless parameter that changes in range $\xi = 0.01 - 100$. The smallest value corresponds to the huge second-order hyperelasticity.

Thus, we have at least one undefined principal parameter. Our further basic calculations were performed for the case of $\xi = 0.01$. It gives us understanding the deformation and forces in the case of the smallest possible interaction with the gold surface.

Another non-acquaintance is that we do not know the degree of initial bacterial bloat. Since Staphylococcus bacteria, under normal conditions, have a stable shape, they inflate like balloons. But the extent of this phenomenon at this stage of research can only be guessed. Here we considered two limits. The first one is the bacterium inflated by only 1% of its unboated state. And the upper limit is the huge inflated bacterium up to 10% of its unboated state.

Taking into account these circumstances the calculations for description of bacterium deformation were performed. The change in the bacterium's shape for different parameter A values is presented in Fig. 3 for an initial inflation of 1% and in Fig. 4 for an initial inflation of 10%. The figure clearly shows that interaction amplification (an increase in the coefficient A) leads to significant deformations. Such a situation arises, in part, when a plasmon is excited on the metal surface.

The comparison of the shapes of the bacterium shown in Figs. 3-4 reveals larger deformations in the case of a small initial inflation. We see a result similar to the deformation of a blown ball when a foot is placed on its top.

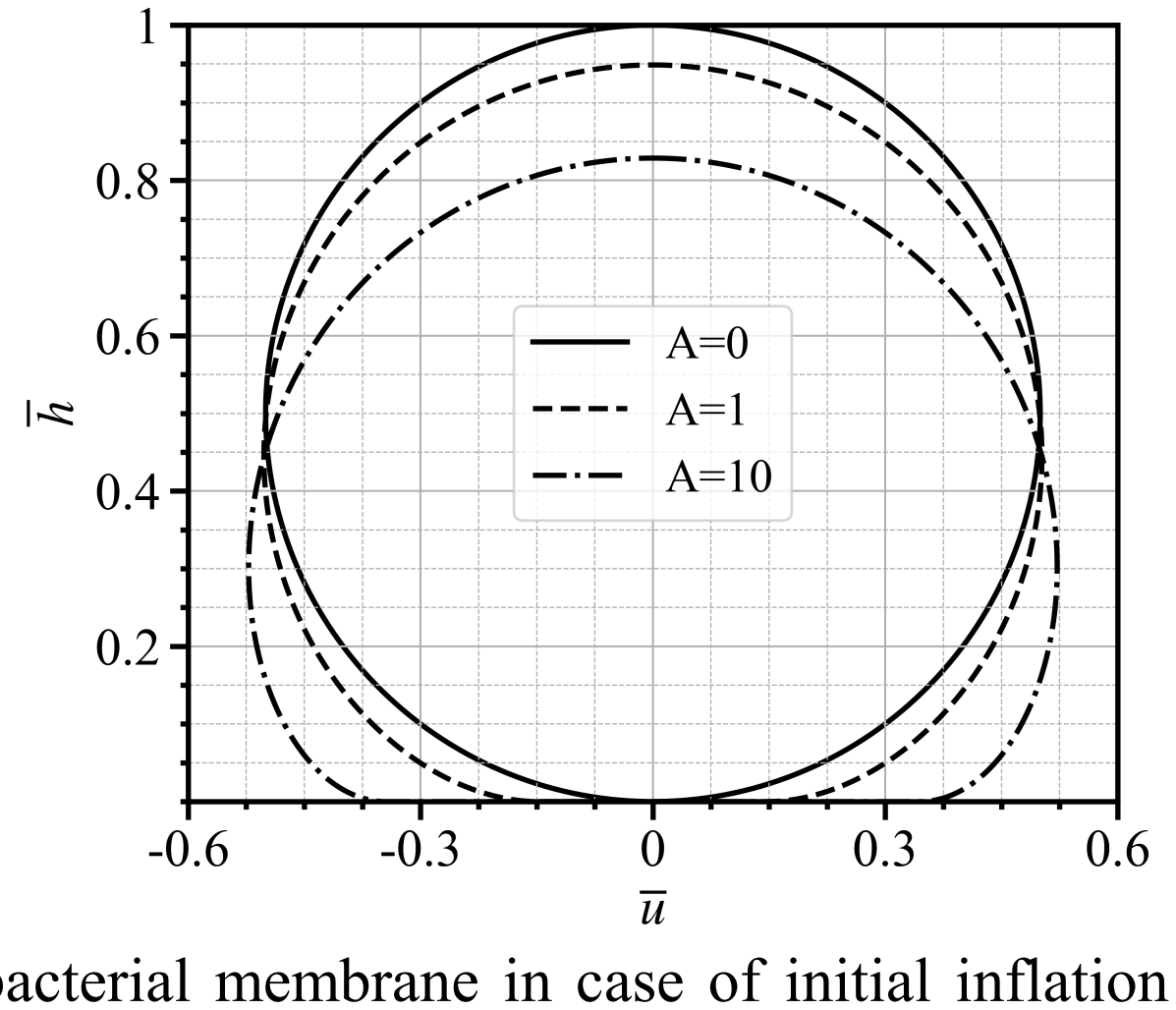

FIG 3. Deformation of bacterial membrane in case of initial inflation by 1% and . The solid black (curve 1) curve represents the shape of free (nondeformed) inflated bacterium. The dashed bleck curve corresponds to the bacterium shape in case of interaction with the surface without plasmon (A=1) and the dash-dot one is the profile of the bacterium in presence of plasmon resonance (A=10).

The more blowed ball we have, the smaller it is deformed. However, we present numerical estimates of the deformation in the vertical direction and the contact area. In the presence of plasmon resonance, for an initial inflation of 1%, the vertical deformation is $0.175d_0$. Moreover for an initial inflation of 10% it is close to $0.075d_0$. The contact area is in the presence of plasmon resonance for initial inflation of 1% and 10%. From these estimates, one can see that the deformation is much larger for small initial inflation and decreases as it increases. It should also be noted that a contact domain appears close to the surface of the metal film, which expands with increasing interaction force. In particular, the domain radius increases by about 1.5 times under conditions of surface plasmon resonance.
More interesting results for additional stresses within the bacterium's shell are shown in Fig. 5 (initial inflation by 1%) and Fig. 6 (initial inflation by 10%). Here, we demonstrate stresses caused by interaction with the metal surface but not by initial inflation itself.

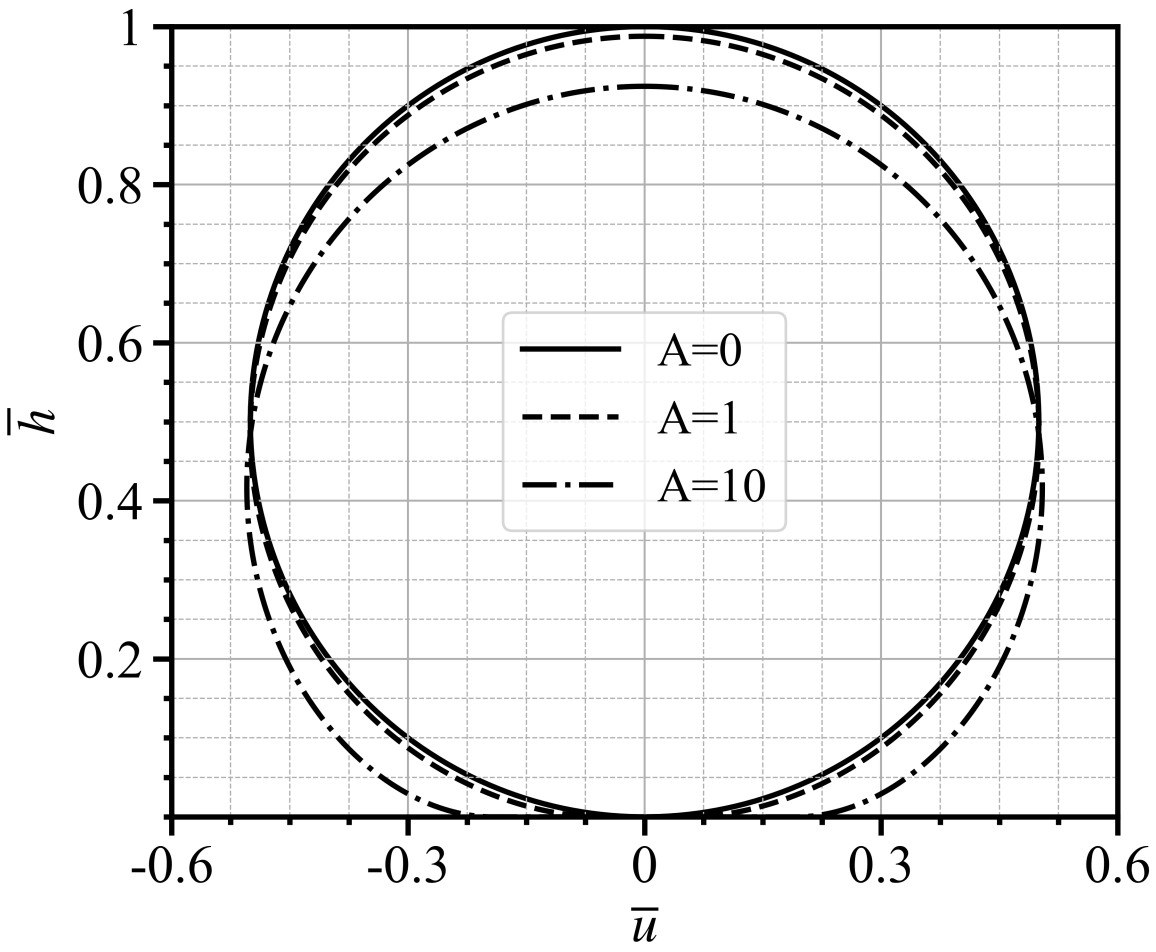

FIG.4. Deformation of bacterial membrane in case of initial inflation by 10% and . The solid black curve represents the shape of free (nondeformed) inflated bacterium. The dashed bleck curve corresponds to the bacterium shape in case of interaction with the surface without plasmon (A=1) and the dash-dot one is the profile of the bacterium in presence of plasmon resonance (A=10).

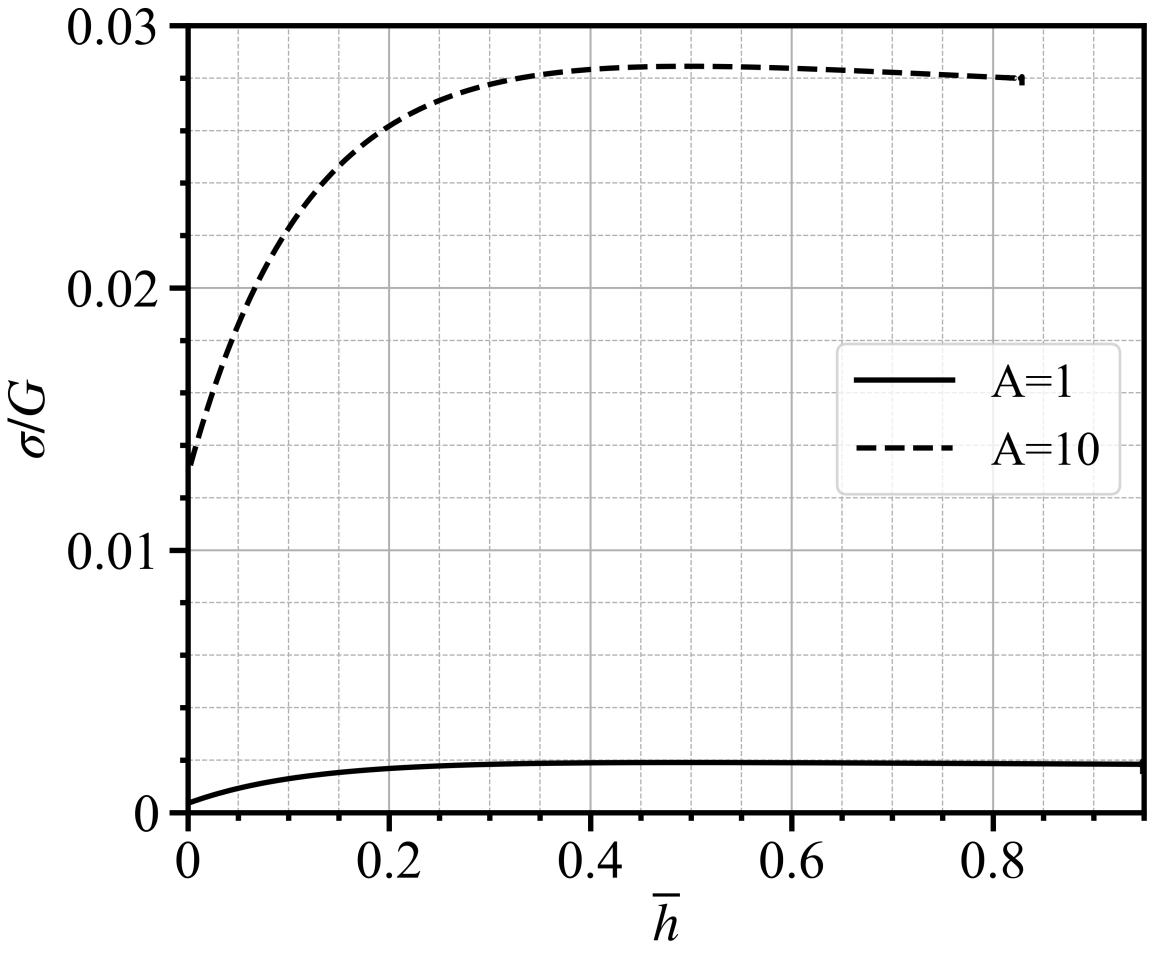

FIG.5. Stress in bacterial shell in case of initial inflation by 1% and . The solid black curve represents the stress profile in a case of the metall surface without plasmon (A=1) and the dash-dot one is the profile in presence of plasmon resonance (A=10).

From the curves of Figs. 5 -6, the more initial inflation, the less additional stress is observed. It is also clear that, in all cases, the presence of plasmon resonance results in dramatic changes in the shell's stress from a flat contact area to the bacterium's height. However, given the bacterium's large nonlinear dielectric properties, the stress values do not cause shell destruction.

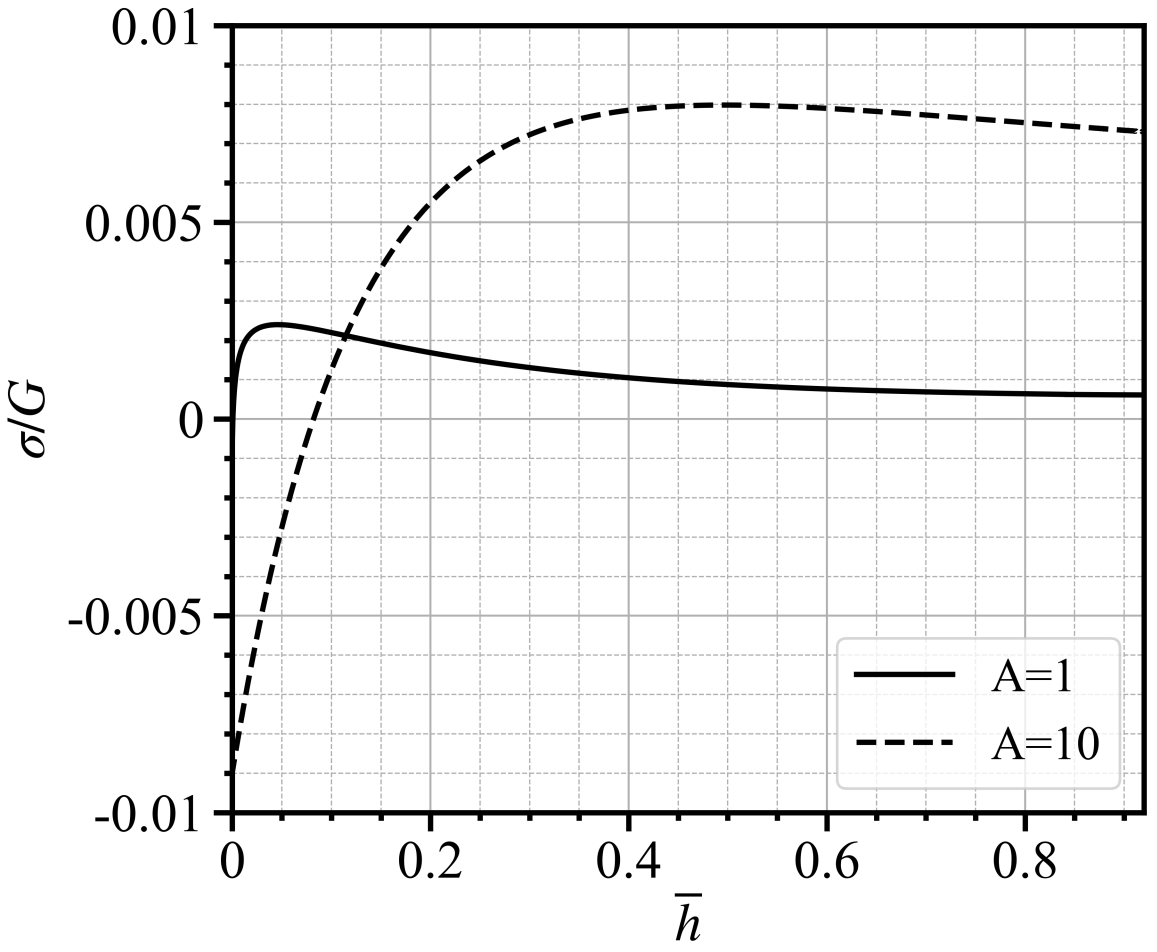

FIG.6. Stress in bacterial membrane in case of initial inflation by 10% and . The solid black curve represents the stress profile in a case of the metall surface without plasmon (A=1) and the dash-dot one is the profile in presence of plasmon resonance (A=10).

The destruction of the bacterial shell, observed earlier [1-3,20,25,26], is clearly associated with the action of ponderomotive forces. These forces were not accounted for in these calculations.

## VIII. DISCUSSION AND CONCLUSIONS

A nonlinear mechanical model of the bacterial shell was developed to evaluate shape deformation during interaction with a metal surface under varying adsorption potential parameters. Notably, under surface plasmon resonance (SPR) conditions—where surface plasmons are excited along the metal interface—the contact domain area increases significantly. This finding provides a mechanical basis for the enhanced antibacterial activity observed on nanostructured gold surfaces under SPR conditions [20].

The antibacterial properties of nanostructured surfaces are linked to the local-field enhancement effect. These properties increase with surface roughness. Namely, local-field gradients near the surface give rise to ponderomotive forces acting on the bacterial membrane. Surface roughness increases local field gradients. The mechanism of action for the nanostructured gold surface can be described as follows: the bacterium adsorbs onto a gold film, which, at the nanoscale, consists of a dense array of "hemispheres" on a flat substrate (Fig. 7). Interaction with these nano-hemispheres induces local-field enhancement, resulting in "hot spots" characterized by high local-field gradients [24]. These gradients generate ponderomotive forces that can damage or destroy molecules within the bacterial membrane.

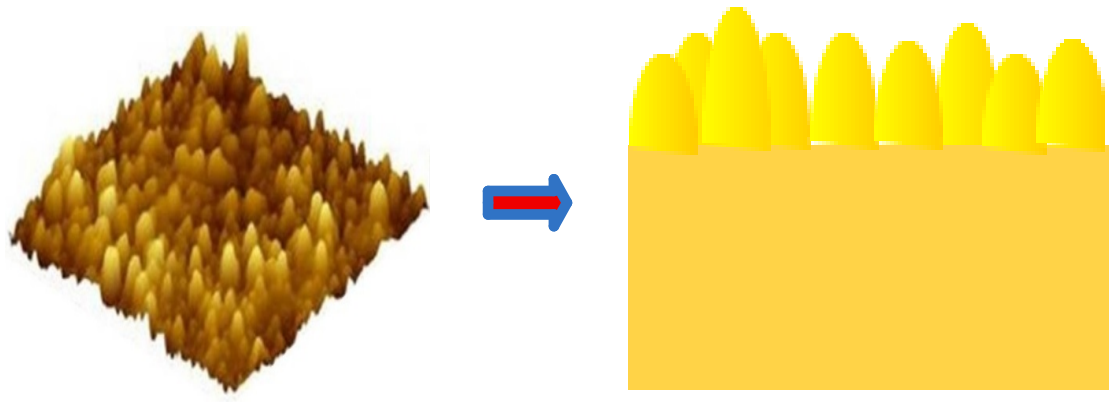

FIG.7. The modelling of real surface of a gold as set of semi-spherical golden elements on the ideal flat surface

According to the modelling results obtained in this work, the excitation of surface plasmons increases the effective interaction area between the bacterial membrane and the nanostructured surface. Consequently, the antibacterial efficacy of the gold surface is greatly amplified under SPR conditions, aligning with experimental observations [20]. The results of previous studies [20] show that plasmon resonance reduces bacterial infectivity through two main pathways:

i. Direct action: Increasing the local-field enhancement effect leads to higher intensity within the hot spots, thereby enhancing the ponderomotive forces acting on the bacterial membrane.

ii. Indirect action: Expanding the effective area of interaction increases the number of "hemispheres" in contact with the bacterial membrane. This expansion increases the cumulative magnitude of the ponderomotive forces.

Here, we did not approximate ponderomotive forces and their action on the bacterial membrane. However, we estimate the interaction with an ideally flat metal surface, both with and without surface plasmon excitation. However, the results obtained provide information on the contact area between the bacterial membrane and the metal surface.

Both pathways increase the efficiency of the ponderomotive forces, which damage the bacterial membrane to the point of destruction, and are tested for an ideally flat surface. Although ponderomotive forces were not applied in this model and can be the subject of further investigation, the results provide a clear understanding of a large increase in membrane deformation, membrane stresses, and the contact area upon plasmon excitation. In the case of an ideally flat surface and a quasistatic interaction, it appears the bacterium will remain stable, although highly deformed. Further studies may reveal new details. In real cases, bacteria move.

For instance, during the flow of a bacterial suspension across the surface, mechanical forces—such as friction—can cause further membrane damage and a subsequent loss of bacterial infectivity [20].
Admittedly, certain physical parameters of the bacterium remain unknown. Present studies suggested calculations for a range of unknown nonlinear physical parameters. These calculations provide a qualitative understanding, which, when synthesised with the experimental data, may solve the inverse problem of determining physical properties from the experimental observation of the deformation process of the bacterial membrane. For instance, the nonlinearity parameter could be approximated by monitoring bacterial destruction experimentally. Furthermore, while the current model provides a foundational framework, further refinement is required to incorporate the dielectric properties of bacterial components and to more accurately assess interaction potentials. These aspects will be the focus of future studies.